\documentclass[lettersize,journal]{IEEEtran}
\usepackage{amsmath,amsfonts}
\usepackage{algorithmic}
\usepackage{algorithm}
\usepackage{array}
\usepackage[caption=false,font=normalsize,labelfont=sf,textfont=sf]{subfig}
\usepackage{textcomp}
\usepackage{stfloats}
\usepackage{url}
\usepackage{verbatim}
\usepackage{graphicx}
\usepackage{cite}
\usepackage{makecell}
\usepackage{textcomp}
\usepackage{array} 
\usepackage{longtable}
\usepackage{booktabs}
\usepackage{float}
\begin{document}

\title{Blockchain technology research and application: a systematic literature review and future trends}

\author{Min An, \IEEEmembership{Graduate Student Member, ~IEEE}, Qiyuan Fan, Hao Yu and Haiyang Zhao
\thanks{This work was supported by the National Natural Science Foundation of China under Grant No. 61862063, 61502413, 61262025; the National Social Science Foundation of China under Grant No. 18BJL104; the Science Foundation of Young and Middle-aged Academic and Technical Leaders of Yunnan under Grant No. 202205AC160040; the Science Foundation of Yunnan Jinzhi Expert Workstation under Grant No. 202205AF150006; the Open Foundation of Yunnan Key Laboratory of Software Engineering under Grant No. 2020SE301; the Science Foundation of "Knowledge-driven intelligent software engineering innovation team".}
\thanks{Min An is with the School of Software and the Yunnan Key Laboratory of Software Engineering, Yunnan University, Kunming 650091, China (e-mail: kirinmin@hotmail.com; 2956384327@qq.com).}
\thanks{Qiyuan Fan, Hao Yu and Haiyang Zhao are with the School of Software, Yunnan University, Kunming 650091, China.}}

\maketitle

\begin{abstract}
Blockchain, as the basis for cryptocurrencies, has received extensive attentions recently. Blockchain serves as an immutable distributed ledger technology which allows transactions to be carried out credibly in a decentralized environment. Blockchain-based applications are springing up, covering numerous fields including financial services, reputation system and Internet of Things (IoT), and so on. However, there are still many challenges of blockchain technology such as scalability, security and other issues waiting to be overcome. This article provides a comprehensive overview of blockchain technology and its applications. We begin with a summary of the development of blockchain, and then give an overview of the blockchain architecture and a systematic review of the research and application of blockchain technology in different fields from the perspective of academic research and industry technology. Furthermore, technical challenges and recent developments are also briefly listed. We also looked at the possible future trends of blockchain.
\end{abstract}


\section{Introduction}
\label{sec:introduction}
\IEEEPARstart{A}{s} an emerging technology, blockchain (BC) has been widely used in many fields, such as manufacturing, logistics and transportation, electronic transactions, intelligent transportation, energy/utilities, healthcare, etc.[1]–[8].

Blockchain is a distributed ledger technology that records and shares every transaction that occurs in the network of users. Nowadays cryptocurrency has become a buzzword in both industry and academia. Digital currencies are only one way to use blockchain. Other evolving applications can include online voting, medical records, insurance policies, property and real estate records, copyrights and licenses, and supply chain tracking [9]. They can also include smart contracts, where payouts between the contracted parties are embedded in the blockchain and automatically execute when contractual conditions have been met. Worldwide spending on blockchain solutions is expected to grow from 4.5 billion U.S. dollars in 2020 to an estimated 19 billion U.S. dollars by 2024 [10]. It can be seen that the value and potential development of blockchain technology is because it is widely used in many fields.

Blockchain was first proposed in 2008 and implemented in 2009[11]. The blockchain is essentially a distributed public ledger with the nature of a key-value database, and all transaction data is permanently recorded in a one-way chain list through asymmetric encryption technology and distributed consensus technology. As new blocks are added and the length of the chain increases, so does the cost of tampering or attack, so the blockchain is also more secure. Blockchain has key characteristics such as decentralization, persistence, anonymity and auditability, which are the key to the continuous popularity of blockchain technology, and the reason why blockchain greatly reduces costs and improves efficiency. Blockchain technology is the representative research of decentralized distributed systems in recent years, but it is also constrained by distributed theory.

Blockchain is a distributed digital architecture system in edge networks, first proposed by Scott Stornetta in 1991[12]. Blockchain is based on a peer-to-peer decentralized network, that is, each blockchain node is a peer-to-peer relationship, and the entire network is a decentralized distributed structure. All network nodes can collect transactions and record them into blocks, and the reward mechanism of the blockchain network can encourage each node in the network to use a unified consensus algorithm to compete for blocks. And eventually form a chain structure composed of blocks stored in each node, so that the blockchain nodes do not need to trust each other. 
\subsection{The state of blockchain research}
With the development of technologies such as the Internet and sensors, the Internet of Things has been continuously applied in various industries, resulting in a large number of new technologies, new products and new applications, and the privacy protection of all intelligent information communication has become very important, thus triggering researchers to innovate and design blockchain technology. Most of the research on blockchain technology mainly focuses on the security mechanism of blockchain and the blocking rate, in which the blocking rate represents the rate at which blockchain network mining generates blocks. This section will introduce the research status of blockchain technology from these two aspects respectively.

\subsubsection{Security mechanism of blockchain}
Blockchain technology mainly uses consensus algorithms to ensure the security and immutable block data, among which, Proof of Work (PoW) is a common consensus protocol in the blockchain system, but when the computing power of the attacker is greater than half of the computing power of the blockchain network, The system will be subject to a double flower attack or 51\% attack by the attacker, which undoubtedly poses a great threat to the blockchain. X.Yang et al. [13] proposed a technique that combines the historically weighted information of miners with the total computational difficulty to mitigate the 51\% attack problem, based on which the cost of traditional attacks increases by two orders of magnitude. J.Bae et al. [14] proposed a random mining group selection technique to reduce the probability of successful double flower attack. This method divides miners into several groups and provides mining opportunities for randomly selected groups. In addition to the optimization based on traditional consensus algorithms to resist attacks, some researchers have also developed some novel consensus algorithms. 

For example, A. Dorri et al. [15] proposed a fast and scalable consensus algorithm named tree-chain. Compared with the existing hash function verification method, Tree chain is a leader selection consensus algorithm. It changes from a linear blockchain structure to a tree-structured blockchain, where each branch is managed by a specific verifier, each verifier is periodically assigned to a range of random consensus codes that match a specific pattern of the most significant bits output by the hash function, and each verifier is then responsible for the transactions it falls within the consensus code assigned to the verifier. Since transactions are deterministic assigned to verifiers based on random consensus codes, tree chains eliminate the significant inefficiencies of traditional blockchains in verifier selection, such as PoW. A. Kumar et al. [16] propose A novel consensus algorithm for public blockchains that shards miners based on their performance. After miner sharding is completed, the best miner from each shard is selected to form a super miner shard, and then a miner is randomly selected from the super shard as the winning miner for mining the next block in the blockchain network, for sharding, the performance history of miners will be maintained in each miner, and resharding will be carried out regularly to bring fairness to the system. This performance-based consensus algorithm ensures more fairness, avoids hunger, improves trust among miners and the overall performance of the blockchain network.

In fact, blockchain technology, due to its data traceability and integrity, can effectively solve the current world's complex distributed system of distributed management and data tampering and other major problems, but the blockchain technology itself can not solve the trust issues related to the data itself, therefore, reputation management has become an important research direction of blockchain technology. S. Malik et al. [17] proposed a three-tier trust management framework named TrustChain, which uses alliance blockchain to track interactions among supply chain participants and dynamically assign trust and reputation scores based on these interactions. Therefore, a reputation model based on this framework can be used to evaluate the quality of goods. And the credibility of entities based on multiple observations of supply chain events. In addition to the research on the security mechanism of the blockchain technology itself, some researchers have studied the deployment cost of the blockchain. For example, P. Frauenthaler et al. [18] introduced a novel relay scheme, considering that the current blockchain relay scheme requires the target blockchain to immediately verify the size of each relay. This results in high operational costs when deploying these relays between Ethereum-based blockchains, and the computational cost of verifying block headers on-chain is high. To overcome these limitations, the scheme uses an on-demand verification model combined with economic incentives to reduce the operational cost of relays between Ethereum-based blockchains by up to 92\%, through which decentralized interoperability between blockchains becomes feasible. In addition, there are also studies focusing on the tamper-proof and security issues of distributed federated learning systems. The model parameters uploaded by terminals are authenticated and shared through blockchain technology, so as to avoid privacy disclosure and tampering attacks caused by long communication distances. The research focus is mainly on the mechanism of blockchain combined with federated learning, and it is also considered in the federated learning scenario based on blockchain technology. The relationship between blockchain block time and system resources and the performance of federated learning system, and the convergence performance of federated learning is optimized.

\subsubsection{The blocking rate of blockchain}
Due to its decentralized property, blockchain is seen as a promising technology for providing reliable and secure services. However, due to the limited throughput, the current blockchain platform cannot meet the transaction needs in actual use, so researchers have proposed many new solutions. S. Yang et al. [19] improved on the linear structure of traditional blockchain protocols using the Directed Acyclic Graph (DAG) structure, in which blocks are organized by level and width, which produces a compact DAG structure (CoDAG). In order to make CoDAG more efficient and secure, algorithms and protocols have also been designed to appropriately place newly generated blocks, which improve security and transaction verification time compared to traditional blockchain protocols, and enjoy the consistency and activity characteristics of blockchain. However, due to the change of the linear structure of the blockchain, the data tracing and verification algorithms have been changed, and under extreme conditions, once there are few blocks associated with a block, it will increase the possibility of being tampered with by attacks, so there is a certain risk. K. Wang et al. [20] extended the traditional analysis of the basic tradeoff between throughput and fork rate of a blockchain system and further proposed FastChain, which increases the throughput of a blockchain system by reducing block propagation time. FastChain employs bandwidth-based neighbor selection, where miners disconnect from bandwidth-constrained neighbors. And prefer nodes with higher bandwidth. This method effectively reduces the propagation delay of block authentication and improves the block rate of the blockchain. However, selecting nodes based on the bandwidth of users may cause unfairness in the system for a long time, and because the connections between nodes are constantly changing dynamically, it will increase the maintenance and management costs of the blockchain network. 

As can be seen from the above research, it is difficult to optimize the mechanism of the blockchain itself to increase the blocking rate. In fact, blockchain technology, as a distributed system protocol, is usually deployed in the edge computing network, and many applications of edge computing, such as the Internet of vehicles, industrial Internet, etc., need a distributed coordination mechanism of security and mutual trust. The application of blockchain in Mobile Edge Computing (MEC) system has aroused great interest of researchers, and task unloading is one of the basic problems of MEC, and collaborative unloading can effectively improve the throughput of distributed systems, which is of great significance for blockchain technology. However, most of the existing work in the design and optimization of blockchain and MEC are carried out separately, which will lead to sub-optimal performance, so some researchers have conducted research on blockchain technology in the scenario based on MEC. J. Feng et al. [21] propose a joint optimization framework for blockchain-based MEC systems to achieve an optimal trade-off between MEC system performance and blockchain system performance. Specifically, the optimal tradeoff of energy consumption and latency (final determination time) is achieved by jointly optimizing user association, data rate allocation, block producer scheduling, and compute resource allocation. A. Vera-Rivera et al. [22] propose a blockchain-based service-oriented architecture that allows secure and private task offloading collaboration among edge servers in MEC environments to help alleviate processing saturation in dense networks and improve resource utilization of mobile edge computing systems. The mechanism is based on the Hyperledger-Fabric blockchain, which serves as a distributed interactive gaming environment with advanced encryption capabilities, minimizing potential security and privacy threats. From the above research, it can be seen that the computing offloading through edge computing can effectively reduce the block rate of the blockchain and improve the system resource utilization.
\subsection{Advantages of blockchain technology}
As a distributed data recording system that provides a secure trust mechanism, blockchain technology is of great significance to the development of various fields at present. This technology has the advantages and characteristics of multi-centralization, multi-party maintenance, time series data, smart contract [23], immutable, open consensus, security and trust. These features are described in detail below.

First, blockchain, as a decentralized distributed system, uses a multi-centralized consensus approach to establish trust mechanisms. The verification, accounting, storage, maintenance and transmission of transactions in the blockchain all rely on the distributed system structure. Each blockchain node can distribute the same mathematical problem to select nodes among multiple distributed nodes for the final mining block, and complete the verification and verification of transactions through the mining process. Thus, instead of the traditional method of using third-party trust organizations or institutions to build trust relationships, a decentralized and trusted distributed system is established [24].

In addition, the blockchain is jointly maintained and participated by all blockchain nodes, and its incentive mechanism ensures that all nodes in the distributed system can participate in the verification process of the block, and through the consensus mechanism, select a specific node to add the newly generated block to the blockchain, thus ensuring the stability and security of the system. Therefore, the blockchain network is a distributed consensus and multi-party maintenance system, which is a robust and secure system. Blockchain technology can also guarantee the authenticity and immutability of transactions. Blockchain technology stores transaction information using a time-stamped chain structure, so the transaction information on the chain can be traced. And for any two adjacent blocks in the blockchain, the information of the latter block contains the information of the previous block, in turn recursively, each block contains the information of all the previous blocks, so once a block is tampered with, the data information of the block and all the subsequent blocks must be modified, and the blockchain maintenance time is limited. Therefore, the tampering operation must be completed within a limited time, but in fact, every re-mining calculation needs to pay a huge price, so the data on the blockchain is immutable.

In addition, blockchain technology can create smart contracts between users who do not trust each other [25], guaranteeing the confidentiality of transactions. Blockchain network is an open consensus network, which encrypts the data on the chain through asymmetric encryption technology to ensure data security, and prevents external attacks from tampering through complex consensus algorithms. In fact, taking PoW consensus algorithm as an example, the computing power of external attacks is at least more than half of the total computing power of the blockchain network to succeed. Therefore, it ensures that the data on the chain is not tampered with and forged, so that it has high confidentiality, credibility and security.

According to the above description, blockchain technology has a tamper-proof mechanism and a security and confidentiality mechanism based on encryption, which is of great significance to ensure the safe sharing of data in the edge network.

Our main contributions of this paper are described as follows.
\begin{enumerate}

\item Starting from the structure and technology of the blockchain itself, we outline its role and type, analyze its key characteristics, and then make a brief summary of its consensus algorithm.

\item We systematically discussed the research and application of blockchain technology in various fields from the perspective of academia and industry.

\item We summarize the current challenges and research progress of blockchain technology, and suggest possible future research trends.

\end{enumerate}

The remainder of this paper is structured as follows. Section II briefly introduces the background of blockchain architecture and discusses the related work. In the Section III, we summarize the overview of blockchain technology research in multiple fields from both academic and industrial perspectives, including the research and application of blockchain technology in federated learning, reinforcement learning, cloud edge computing, intelligent transportation, smart grid and IoT. We describe the current challenges about blockchain in Section IV, and introduce the research progress of blockchain technology respectively. Possible future research trends are described in Section V. Section VI concludes this paper.
\section{Blockchain architecture}
\subsection{Blockchain structure}
Blockchain is a sequence of blocks, which holds a complete list of transaction records like conventional public ledger. Figure 1 shows the basic structure of the blockchain, where TX represents a specific transaction on the blockchain. The underlying data structure of the blockchain is shown in Figure 2. The blockchain starts from the genesis block, and the orderly one-way connection in the way of Hash pointer constitutes the whole blockchain, and the validity of transactions on the chain is guaranteed according to the longest chain legal principle and consensus algorithm.

\begin{figure}[h]
  \centering
  \includegraphics[width=0.5\textwidth]{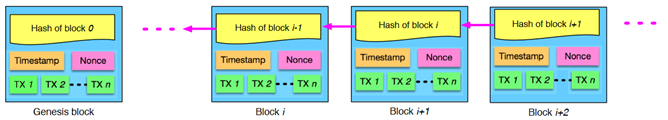}
  \caption{ \centering An example of blockchain which consists of a continuous sequence of blocks.}
  \label{fig1}
\end{figure}

A block on the blockchain is composed of two parts: a block and a block. The block contains data records generated within a certain period of time that cannot be tampered with. Specifically, the block contains information such as the block version, Merkle tree root hash, timestamp, parent block hash, and nonce. A block consists of a transaction counter and all transactions within the block. The maximum number of transactions that a block can contain mainly depends on the size of a single block and the size of each transaction, and the maximum number of transactions essentially represents the throughput performance indicator of the blockchain. The typical digital signature algorithm used in blockchains is the elliptic curve digital signature algorithm (ECDSA) [26].

\begin{figure}[h]
  \centering
  \includegraphics[width=0.5\textwidth]{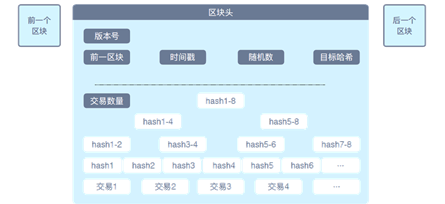}
  \caption{\centering Diagram of the underlying data structure of the blockchain.}
  \label{fig2}
\end{figure}

In addition, in order to enhance the usefulness of blockchain on resource - and power-constrained devices, people propose a new DAG-Structured blockchain, based on directed acyclic graph architecture. In chain-structured blockchain, a new transaction must be validated before attached to the main chain, which is called synchronous consensus. Different from it, tangle adopts an asynchronous consensus, which is more efficient in improving system throughput. 

\begin{figure}[h]
  \centering
  \includegraphics[width=0.5\textwidth]{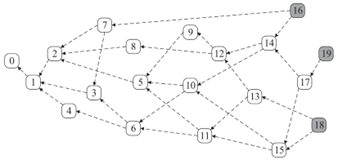}\hspace{0pt}
   \caption{\centering DAG-structured blockchain.}
  \label{fig3}
\end{figure}

As shown in Fig. 3, DAG-structured blockchain is not constrained by the single main chain and forks all the time, the relation among transactions looks like a tangled net. This novel architecture and consensus mechanism can improve network throughput and system response time theoretically. IOTA[27], Byteball[28], and NANO are three representative DAG-structured blockchains.
\subsection{Blockchain type}
Blockchain is broadly categorized into three types: public or permissionless blockchain, private or permissioned blockchain, and consortium blockchain[29]. We compare these three types of blockchain from different perspectives. The comparison is listed in Table 1. 

In a public blockchain, there is no dominant authority and no party has more power than others in the network. Participants can enter and exit at any time according to their wish. Similarly, any participant can validate the transaction due to its public nature. In Bitcoin, for example, miners can validate the transactions and receive Bitcoins as rewards. With a private blockchain, a centralized structure is followed, where a single entity has full power to validate the transactions and make decisions. The private blockchain is more efficient, easy to implement, utilizes fewer energy resources, and is faster compared to the public blockchain. Besides, with the consortium blockchain, not every member has the same permissions. A few members of the blockchain network are assigned certain privileges to validate the new blocks. Other members can also validate but must reach a consensus before implementation. Different consensus algorithms are implemented depending on the requirements and environment. 

\renewcommand\arraystretch{1.5}
\begin{table*}
 \caption{\centering Comparisons among public blockchain, consortium blockchain and private blockchain}
    \centering
   \begin{tabular}{@{}lccc@{}}
     \Xhline{1.2pt}
    \textbf{Property} & \textbf{Public blockchain} & \textbf{Consortium blockchain} & \textbf{Private blockchain} \\
     \Xhline{1.2pt}
    \textbf{Consensus determination} & All miners & Selected set of nodes & One organization \\
    \textbf{Read permission} & Public & Could be public or restricted & Could be public or restricted \\
    \textbf{Immutability} & Nearly impossible to tamper & Could be tampered & Could be tampered \\
    \textbf{Efficiency} & Low & High & High \\
    \textbf{Centralized} & No & Partial & Yes \\
    \textbf{Consensus process} & Permissionless & Permissioned & Permissioned \\
     \Xhline{1.2pt}
  \end{tabular}
 
  \label{tab:blockchain}
\end{table*}

\begin{table*}[!ht]
\caption{\centering Comparison of Smart Contract Platforms}
  \centering
  \begin{tabular}{@{}lccc@{}}
    \toprule
    \textbf{Platform} & \textbf{Language} & \textbf{Consensus Protocols} & \textbf{Permission} \\
    \midrule
     Ethereum & Solidity, Serpent, LLL, Mutan & PoW & Public  \\
        Corda & Java, Kotlin & Raft & Private  \\ 
        Fabric & Java, Golang & PBFT & Private  \\ 
        Rootstock & Solidity & PoW & Public  \\ 
        EOS & Python, JavaScript, Golang and PHP & BFT-DPOS & Public  \\ 
        Stellar & C++ & Stellar Consensus Protocol (SCP) & Consortium, Private  \\
    \bottomrule
  \end{tabular}
  \label{tab:smart-contract}
\end{table*}
Consensus algorithms are the core of blockchain and determine how it will work. It is the critical technology that describes the security and improves the performance of blockchain. A consensus algorithm means an agreement, used in a decentralized network communally to collectively make a decision when it is needed. Its properties include non-repudiation, authentication, decentralized control, transparency, and byzantine fault tolerance[30]. Authors[31], elaborated the five components of the consensus algorithm: (1) block proposal, (2) block validation, (3) information propagation, (4) block finalization, and (5) incentive mechanism. In addition, famous consensus algorithms are such as Proof of Work (PoW), Proof of Skate (PoS), Proof of Existence (PoE), Proof of Authority (PoA), etc. Another term smart contracts [32] are deployed in blockchain as a digital agreement between two or many other parties. The concept of smart contracts was first proposed by Nick Szabo[33] in 1994 and was first implemented in Ethereum. Smart contracts are computer programs that can be consistently executed by a network of mutually distrusting nodes, without the arbitration of a trusted authority. Being embedded in blockchains, smart contracts enable the contractual terms of an agreement to be enforced automati-cally without the intervention of a trusted third party. Based on its pre-defined function, it can store, process information, and write outputs. To prevent tampering, smart contracts are copied to each node in the blockchain. Besides, a smart contract enables transaction traceability in FL as well as irreversibility[34].

Table 2 compares  Ethereum, Fabric, Corda, Stella, Rootstock (RSK) and EOS from the following  aspects such as execution environment, supporting language, Turing completeness, data model, consensus protocols, permission and application.

\subsection{Key features of blockchain}
In summary, block chain has the following characteristics.

Decentralization. In conventional centralized transaction systems, each transaction needs to be validated through the central trusted agency (e.g., the central bank), inevitably resulting to the cost and the performance bottlenecks at the central servers. Contrast to the centralized mode, third party is no longer needed in blockchain. Consensus algorithms in blockchain are used to maintain data consistency in distributed network.

Persistency. Transactions can be validated quickly and invalid transactions would not be admitted by honest miners. It is nearly impossible to delete or rollback transactions once they are included in the blockchain. Blocks that contain invalid transactions could be discovered immediately.

Anonymity. Each user can interact with the blockchain with a generated address, which does not reveal the real identity of the user. Note that blockchain cannot guarantee the perfect privacy preservation due to the intrinsic constraint (details will be discussed in section IV).

Auditability. Bitcoin blockchain stores data about user balances based on the Unspent Transaction Output (UTXO) model [11]: Any transaction has to refer to some previous unspent transactions. Once the current transaction is recorded into the blockchain, the state of those referred unspent transactions switch from unspent to spent. So transactions could be easily verified and tracked.

Credibility. As a distributed system, blockchain adopts a consensus algorithm to ensure the validity of on-chain transactions. Double-spend attacks are prevented using UTXO structures, a unique data structure that ensures that the input for every transaction on the chain has a legitimate source. At the same time, because the Hash value link is used, the modification of on-chain information depends on the contents and calculation order of a series of data in the previous block. The huge cost and uncertain results brought by such tampering ensure that the data on the chain is difficult to tamper with. At the same time, the blockchain consensus mechanism guarantees that the latest blockchain is accurately added to the blockchain in the shortest possible time, and the blockchain information stored by the nodes is consistent and does not fork or even resist malicious attacks.
\subsection{Consensus algorithm of blockchain}
The consensus algorithm of blockchain technology is an important means for decentralized distributed network nodes to reach consensus, so that in the absence of a central node, each node can still efficiently collaborate to record and store transaction data. There are three typical consensus algorithms: Proof of work[35], Proof of authority (PoA)[36], and proof of stake (PoS)[37]. A comparison of common consensus mechanisms is shown in Table 3.

\begin{table*}[!ht]
\caption{Comparison of major consensus mechanisms}
\centering
    \begin{tabular}{cccccc}
   \Xhline{1.2pt}
        ~ & PoW & PoS & DPoS & Raft  \\ \hline
        Application Scenarios & Public blockchain & Permissioned blockchain & Permissioned blockchain & Consortium blockchain  \\
        Degree of decentralization & Fully decentralized & Fully decentralized & Fully decentralized & Semi-decentralized  \\ 
        Accounting node & Full network & Full network & Selected nodes & Leader-based  \\ 
        Response time & About 10 minutes & About 1 minute & About 3 seconds & Second level  \\ 
        Throughput & About 7 TPS & - & About 300 TPS & -  \\ 
        Storage efficiency & Full ledger & Full ledger & Full ledger & Full ledger  \\ 
        fault-tolerant & 50\% & 50\% & 50\% & 50\%  \\ 
        \Xhline{1.2pt}
    \end{tabular}
\end{table*}

Among them, PoW is the most traditional consensus mechanism, and its consensus process is mainly based on the computing power of each node on the blockchain to carry out mining, which is essentially a hash calculation of transactions. Ideally, the node that completes the calculation the fastest has the accounting right of the block and gets the corresponding mining reward. The blockchain network based on the PoW consensus algorithm is completely decentralized, does not need to structure trust institutions, and the block rate of the whole network is completely dependent on the computing power of the network, at the same time, as long as the computing power of the attacker does not exceed half of the computing power of the whole network, transactions can be recorded normally, and cannot be tampered with, so it is a safe and trusted method. However, the method based on computing power will inevitably cause a large amount of computing resources and energy waste in the mining process, and nodes with more computing power will inevitably generate more rewards, and the hash calculation time is long, resulting in a long block cycle. PoS consensus algorithm reduces the waste of computing power compared with PoW consensus algorithm. The algorithm uses virtual resources such as the number of tokens held by a node or token time to characterize the equity of each node, and uses the blockchain node with the highest equity to make the final accounting for the block. Based on the consensus algorithm, the rights and interests of nodes will be depreciated because of the damage to the security of the system, so the authenticity of node information can be guaranteed. 
Although the PoS consensus algorithm reduces the waste of computing power and speeds up the computing rate to a certain extent, with the longer the blockchain exists, some long-standing nodes are likely to have huge rights, resulting in too much concentration of interests, and it is difficult to fairly carry out node block selection. PoA consensus algorithm can effectively reduce the waste of computing power resources and avoid 51\% computing power attacks. The algorithm mainly selects the blockchain node through voting for block accounting. After selecting the accounting node, the transaction will be directly sent to the node for authentication, so there is a centralized problem and low efficiency.

\section{Multi-field application research }
\subsection{Blockchain-based federated learning}
The traditional federated learning architecture collects and aggregates model information of all participants based on a central server, and then the central server merges and updates the model information to the participants. 

This process may lead to the following three problems: 
\begin{enumerate}
\item The central node may be unstable due to the influence of service providers or other computing tasks; 
\item The central node may favor some clients, resulting in unfair system; 
\item If the central node is attacked or malicious, the training of the model will be damaged or the data privacy will be disclosed. 
\end{enumerate}
In addition, the central node is usually far away from the terminal, and it is easy to be eavesdropped or tampered with by attacks during the interaction process. In order to solve the above problems caused by the central structure, decentralization through the edge network is an effective method. In order to ensure the security and privacy of parameter aggregation in the edge network, federated learning combined with the blockchain technology in the edge network shows its advantages of security and convenient coordination. In fact, at present, many studies have taken blockchain technology as the infrastructure of federated learning to realize the task of model aggregation of federated learning through blockchain technology, and the incentive mechanism in blockchain also provides technical solutions to improve the enthusiasm of participants to participate in the training of federated learning model.

Blockchain as a decentralized distributed storage architecture, with its asymmetric encryption and consensus algorithm to ensure that it can not be tampered with and cannot be forged, so as to ensure the authenticity of the parameters of the federated learning process. Although the traditional federated learning process is decentralized through blockchain technology, the security of the federated learning model storage and update process has been greatly improved, but the introduction of blockchain also brings many new problems and challenges, for example, the transactions of the blockchain network are randomly arrived and randomly blocked. This will inevitably cause a large waiting delay for each periodic federated learning process; Moreover, because of the dynamic change of network resources, it is difficult to predict the block time of each round, so it is difficult to adjust and control the convergence time of federated learning. In addition, blockchain block is usually based on computing power or equity block, compared to the previous centralized structure, the block process will consume more delay and energy consumption.

Once the federated learning process has decentralized the centralized structure using the edge network, it is difficult for the participants to coordinate and interact safely and effectively under the condition of mutual distrust. Generally, the distributed system builds trust based on trusted third-party institutions or organizations to provide trust authentication and cryptographic interaction. However, in the absence of trusted third-party institutions or organizations, it is difficult for distributed systems to directly establish trust relationships. Therefore, blockchain technology has important implications for establishing a secure and trusted system for federated learning[38]. Figure 4 shows the architecture of a federated learning network based on blockchain.

C. Xu et al. [39] proposed an asynchronous federated learning framework with dynamic scaling factor based on blockchain. The framework addresses the issue of trust between devices primarily through blockchain, and at the same time, proposes new dynamic scaling factors to help improve FL efficiency and accuracy. The framework mitigates the impact of low-performance devices while being just as efficient as traditional FL and has the added benefit of alleviating trust issues between IoT devices. H. Jin et al. [40] considered the problem that the existing blockchain-based federated learning solutions perform poorly when the data in the blockchain-based federated learning (BFL) cluster is sparse. A direct solution is to collect as many devices as possible to establish a large BFL cluster. These devices may be located in geographically distant regions and far apart, which will result in high communication latency, which will result in system inefficiencies for BFL due to frequent communication in blockchain consensus. So they propose CFL, a cross-cluster FL system facilitated by cross-chain technology that divides large clusters into multiple smaller clusters, each within its own geographic area and organized by BFL. CFL connects multiple BFL clusters, where only a few aggregated updates are transmitted over long distances across clusters, thereby improving system efficiency. The design of CFL focuses on cross-chain consensus protocol to ensure the safe exchange of model updates between clusters, while the upload parameters of federated learning are still blocked based on the traditional consensus mode within the chain. Therefore, further optimization of the block delay within the chain can effectively improve the efficiency of the system. It can be seen that the federated learning framework based on blockchain technology has become an important direction for the development and research of federated learning. However, in addition to the blockchain mechanism to ensure the security of data, the throughput rate of blockchain itself is limited, which will limit the efficiency of federated learning. In the current research, more consideration is given to optimizing the model update process of federated learning. There is no consideration for the optimization of the blockchain mechanism as well as the mechanism of the system itself.

\begin{figure}[h]
  \centering
  \includegraphics[width=0.5\textwidth]{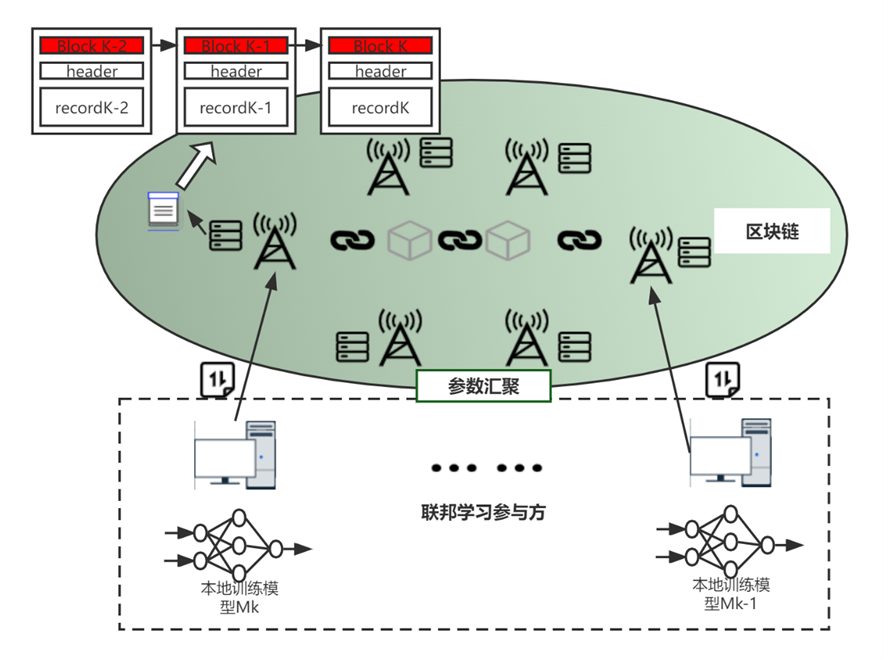}
   \caption{\centering Federated learning network architecture based on blockchain.}
  \label{fig4}
\end{figure}
\subsection{Blockchain with reinforcement learning}
Currently, reinforcement learning is mainly used to optimize the performance of blockchains. Currently, most IIoT applications rely on centralized servers for data processing and transmission, which exposes data to security risks as well as high operational costs and latency. Therefore, data security and efficiency become key issues for IIoT.

To address the above issues, blockchain is widely recognized as a promising solution for building a secure and efficient data storage/processing/sharing environment in the IIoT. Blockchain was originally used as a Peer-to-Peer (P2P) ledger for Bitcoin economic transactions, which can ensure data security and efficiency by enabling anonymous trusted transactions and removing various intermediaries. Despite the significant benefits of blockchain technology, traditional blockchain systems struggle to provide the scalability needed to meet the high transaction throughput demands of the IIoT. In fact, scalability has become a key issue for blockchain to be used as a common platform for different services and applications. Bitcoin, as the first blockchain-based cryptocurrency, can only confirm about 3-4 transactions per second on average, and Ethereum has increased throughput to about 14 TPS, which is still not enough to handle high-frequency trading scenarios such as the Industrial Internet of Things.

In recent years, many teams have been working to achieve a universal, scalable, and deployable blockchain platform. It is mainly divided into two kinds according to different optimization angles. One is on-chain optimization schemes such as adjusting block sizes and intervals (e.g., Bitcoin Cash), improving the block out process (e.g., Bitcoin-NG), proposing new consensus mechanisms such as proof of equity (PoS), proof of commission risk (DPoS), and actual Byzantine fault tolerance (PBFT)(e.g., Cardano and EOS). Sharding technology (e.g. Zilliqa). One is the off-chain optimization scheme, which aims to reduce redundancy on the main chain using side chains, multi-chains (such as Cosmos and AION), lightning networks, payment channels (such as Raiden networks and TeeChan), etc.

In order to deal with the dynamic and high-dimensional characteristics of IIoT systems, M. Liu et al. [41] designed a DRL-based algorithm to dynamically select/adjust block producers, consensus algorithms, block sizes, and block intervals to improve performance. In it, a performance optimization framework for blockchain IIo T systems is proposed for the first time to improve the performance of data security and efficiency, which considers the four-fold tradeoffs of scalability, decentralization, security and latency. M. Liu et al. [42] proposed a new blockchain vehicle networking performance optimization framework based on deep reinforcement learning (DRL) to maximize transaction throughput while ensuring the decentralization, delay and security of the underlying blockchain system. In this framework, we first analyze the performance of blockchain systems in terms of scalability, decentralization, latency, and security. DRL technology is used to select block producers and adjust block size and block interval to adapt to the dynamic changes of vehicle networking scenarios. Simulation results show that our proposed framework can effectively improve the throughput of blockchain-enabled vehicle-connected systems without affecting other characteristics.
\subsection{Blockchain and cloud-edge computing}
Blockchain is considered to be the prototype of the next generation of cloud computing, and the combination of blockchain and cloud computing is a hot research area. In November 2015, Microsoft proposed the concept of blockchain as a service (BaaS), deploying blockchain as an application service in the cloud. This is similar to the software-as-a-service (SaaS) model, where users can interact with different technologies in a low-risk environment provided by the Azure cloud platform. Tencent FiT released a white paper on blockchain solutions in April 2017, building an enterprise-class blockchain infrastructure platform, using blockchain as the underlying underlying technology, adding modules such as operation monitoring and user management to achieve effective supervision of the system, so as to achieve a safe, reliable and flexible blockchain cloud service. Users can flexibly and autonomously implement various blockchain applications based on the basic framework provided by the platform.

The limited computing and storage resources of edge servers not only need to provide support for relevant application services, but also need to cache data frequently accessed by users, so deploying blockchain should save server resources as much as possible. In a cloud computing environment, multiple copies of data backup are placed in multiple data centers to cope with the access requests of a large number of users in different locations, but the placement of data copies should take into account the storage cost of the data center, the access delay of users, and the transmission cost between servers. The blockchain system oriented to the edge computing field should optimize and improve the blockchain to reduce the overhead of block transmission and storage. Farhadi et al. [43] proposed a blockchain system supported by fog computing architecture, but each fog node needs to save complete data, resulting in huge network transmission costs and node storage costs. Ayoade et al. [44] proposed a decentralized data management system, which uses smart contracts to manage access rights and save the hash of data in blocks, reducing the storage cost of blocks, but storing complete data in additional devices still increases the storage cost. Huang et al. [45] proposed an edge-computing oriented blockchain system, which realizes the optimal storage strategy on edge devices including mobile devices, improves the proof-of-stake mechanism to meet the edge environment, and reduces the storage cost of nodes. However, the device storing blocks is an intelligent device with limited resources and high mobility.

The edge cloud environment of cloud edge aggregate computing meets the distributed requirements of blockchain deployment, and the blockchain can be deployed in the edge cloud to ensure the security and reliability of data uploaded at the edge end. However, the shortcomings of blockchain storage occupation and resource consumption still hinder the deployment of blockchain in the edge cloud, and the main task of the edge server is still to process the task of uploading edge devices to achieve real-time effect, and the blockchain deployed in the edge server should save server resources as much as possible. Obviously, this poses a challenge to the deployment of blockchain, and researchers have designed a lightweight blockchain LBlockchainE that is suitable for the edge cloud in the cloud edge converged computing environment.
\subsection{Blockchain and intelligent transportation}
Blockchain applications in smart transportation are equally widespread. Based on the immutability and traceability of blockchain, some parking reservation solutions can be designed in combination with the reputation mechanism. Currently, the application of blockchain technology in the Internet of Vehicles (IoV) has received some attention. In order to solve the problems in iot, some researchers have proposed different combinations of iot and blockchain solutions (such as privacy protection, vehicle life cycle, vehicle supply chain, vehicle edge computing, electronic toll collection).

Intelligent Transportation System (ITS) is critical to cope with traffic events, e.g., traffic jams and accidents, and provide services for personal traveling. Although some researches have investigated the integration of blockchain and ITS, they mainly focus on data sharing, energy delivery, trust management, blockchain-enabled crowdsensing, and blockchain network architecture. However, to the best of our knowledge, existing researches ignore the blockchain safety, the brought latency by blockchain and the trade-off between these two metrics.

\begin{figure}[h]
  \centering
  \includegraphics[width=0.5\textwidth]{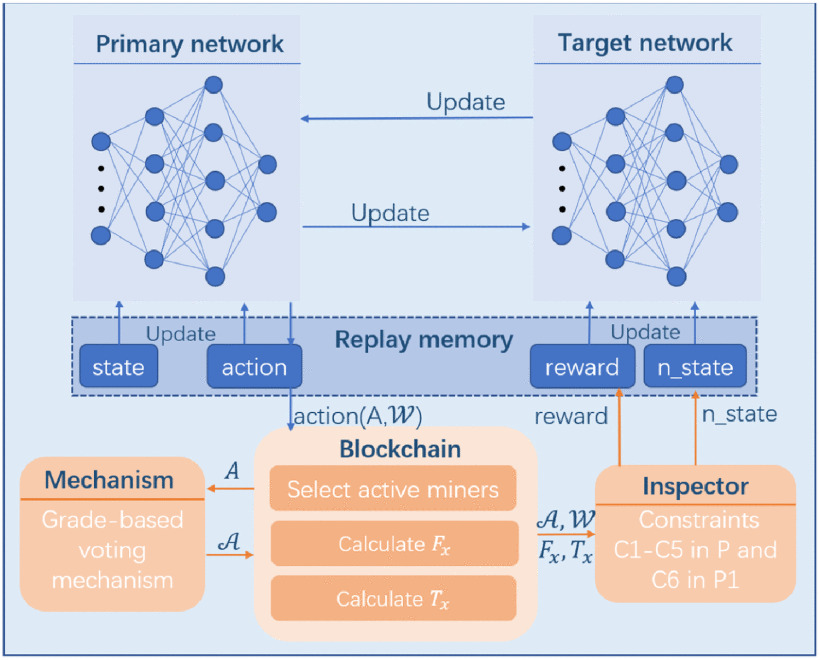}
   \caption{\centering The framework of DRL-based algorithm for ITS transaction selection[46].}
  \label{fig5}
\end{figure}
When blockchain and ITS are combined, for transaction selection, we need to consider the data size, waiting time in the transaction pool and blockchain environment. However, it is rather challenging to select suitable active miners from RSUs, since the reliability and computing power of RSUs need to be evaluated. To solve the problem, a DRL-based algorithm is proposed to select transactions and active miners, and the corresponding framework is shown in Fig. 4. Z. Ning et al. [46] put forward a secure, efficient and distributed ITS system, and formulated it as a multi-objective optimization problem, i.e., minimizing the system latency, maximizing the data safety and user utility. In order to solve the formulated problem, we decomposed it into two subproblems, and proposed two corresponding algorithms. The DRL-based algorithm can make a satisfied trade-off between blockchain security and latency, and the DIADEM algorithm is able to choose task computation modes for vehicles in a distributed way. Extensive experiments demonstrate the effectiveness of our algorithms, i.e., the DRL-based algorithm can reach higher blockchain safety and lower blockchain latency, and the DIADEM algorithm can obtain larger social welfare than benchmark methods.
\subsection{Blockchain and power system}
Blockchain promises to change the way we execute global value transactions. Therefore, it is very important to explore the application of this new technology in the field of energy. This part mainly expounds the application of blockchain in the energy field from the perspective of energy trading and power grid operation. We see blockchain as a distributed system that can provide trust between different and independent parties. It allows the creation of distributed peer-to-peer (P2P) networks where untrusted members can interact in a verifiable way without a trusted intermediary. With blockchain, transactions can be managed in ledger form, making microgrids more powerful. These transactions may include electricity transactions, currency transactions or even records of the flow of electricity in the network.

At present, electricity trading has been able to realize online trading, but this trading method is still in a relatively preliminary stage, that is, the use of centralized central database to store and process electricity consumption data and transaction data, this method may encounter network attacks in the Internet, and at the same time, the data center operator as an intermediary between the seller and the buyer in electricity trading. There is no guarantee of trusted transactions, so blockchain technology could enable new power trading systems to solve the above problems.

Moreover, with the continuous expansion of the scale of power and the gradual opening of the power market, the operational stability and security of the power grid tend to decline, and the power grid is heavily dependent on external power, which is difficult to meet the power supply needs of diverse users, and the development of new energy, the problem of new energy consumption and utilization is more and more serious. In view of this, an environmentally friendly, efficient and flexible micro-grid system integrating power generation, distribution and sale has emerged. The development of microgrid helps to solve the problem of new energy consumption, and its deployment at the receiving end makes it easy to meet the diversified needs of the receiving end users. Blockchain is known as the "next generation Internet", as a decentralized, low trust cost, information cannot be tampered with distributed ledger system, its combination with distributed microgrid system has become a research trend.

During the construction of the new power system, due to the access of various subjects such as micro-grid, distributed generation unit and new energy power system, as well as the improvement of the intelligence of the new power system, the power dispatching data presents the characteristics of decentralized data sources, large data volume and high data transmission frequency. The interactive ability of data transmission and processing in power system, especially in power dispatching system, is required to be higher. At present, the power dispatching data mostly adopts one-to-many transmission mode, and the higher dispatching authority only interacts with the lower dispatching authority within the jurisdiction. However, on the one hand, this kind of interaction brings great pressure to the uplink data transmission channel, and it is not easy to maintain and manage. On the other hand, the efficiency of data sharing and interaction between dispatching agencies at all levels, dispatching systems and other systems is low, and it is impossible to synchronize power dispatching data in real time and dynamics. Moreover, the risk of power dispatching data being tampered during transmission cannot effectively ensure the security of power dispatching data. Thus, it brings some difficulties to the development of new scheduling services such as load storage and load scheduling.

The power dispatching data processing method based on blockchain first verifies the identity information of the power dispatching system. When the identity information verification passes, a secure channel between the power dispatching system and the blockchain is established for the data interaction between the power dispatching system and the blockchain, which can effectively prevent unauthorized users from obtaining power dispatching data in the interaction process. Improve the reliability and confidentiality of data interaction, and then establish a safe channel to realize the power dispatching data of the power dispatching system and the data processing results of the blockchain, and carry out data interaction between the power dispatching system and the blockchain to realize the safe interaction of the whole process of power dispatching data up-chain, on-chain processing and reading, and can synchronize the power dispatching data in real time and dynamically. Effectively reduce the difficulty of maintenance and management of a large amount of power dispatching data.

In addition to that, Blockchain technology are also widely used in Power flow[47]–[49], Emission Reduction[50]–[52], Energy Markets[53]–[75], Batteries[76], [77], Demand Response[78]–[84], Electric Vehicles[85]–[94], Security and Privacy[95]–[106], and Other[107]–[114].
\subsection{Blockchain and IoT}
The integration of the Internet of Things and industry is an important means to promote industrial automation and information. IIoT helps reduce errors, reduce costs, improve efficiency and enhance security in manufacturing and industrial processes, enabling a higher level of integrity, availability and scalability in the industrial sector. However, security attacks and failures could cause huge headaches for the global iot network. For example, central data centers are vulnerable to single point failures and malicious attacks such as DDoS, Sybil attacks, etc. In addition, there is a risk of leakage of sensor data stored in data centers. In addition, communication between iot devices may be subject to data interception, and the credibility of the collected data cannot be guaranteed.

In recent years, with the emergence of blockchain, the idea of combining blockchain with the Internet of Things has gained widespread attention [115]–[117]. Leveraging the characteristics of blockchain's tamper-proof and decentralized consensus mechanism, there is an opportunity to address the security issues in the IIoT systems described above.

There are some existing research on this topic, for example, O. Novo[115] proposes an access control system based on the blockchain technology to manage IoT devices. However, the system is not fully built on a distributed architecture because of the usage of the central management hub. Once the management hub is failed or attacked, IoT devices connected to it become unavailable. Z. Li et al.[102] exploit the consortium blockchain technology to propose a secure energy trading system. But they do not consider privacy issues, such as the sensitive data disclosure risk, and thus it cannot guarantee sensitive data confidentiality. The aforementioned systems all adopt chain-structured blockchains in IoT systems, which are overloaded for power-constrained IoT devices. Z. Xiong et al. [118] introduce edge computing for mobile blockchain applications and present a Stackelberg game model for efficient edge resource management for mobile blockchain. They reduce computational requirements of mobile devices by leveraging edge computing. In addition, there are some other challenges that also brought in the meantime when introducing the novel design of blockchain into IIoT systems.
\section{Challenges \& Recent advances}
Although blockchain has been widely used in many fields and achieved great success, and based on its huge potential is still being studied. But at the same time, it also faces many challenges that limit its further application. Below, we list some of the key challenges and recent research developments from the perspective of blockchain application limitations.
\subsection{Performance optimization}
Performance optimization is mainly divided into two perspectives, one is the optimization for specific scenarios, and the other is the optimization for the overall transaction performance of the blockchain. In resource-constrained scenarios, the computing and storage resources of each node are limited, but the proof-of-work based consensus algorithm of blockchain itself needs to consume a lot of computing resources to compete for bookkeeping rights, which is not applicable in cloud-edge computing scenarios. At this time, the low-energy characteristics of the proof-of-stake mechanism can be used in combination with lightweight blockchain to determine the ownership of accounting rights through a small amount of competitive computing and node resources. The optimization of the overall transaction performance for the blockchain is mainly because of the blockchain expansion problem, with the increase of transaction volume, it will inevitably bring huge storage costs, and increase the blockchain transaction delay. The nodes participating in the consensus must store all transactions to verify the legitimacy of the transaction on the blockchain, in addition, due to the block size and block time limits, the transactions that the blockchain can process in a certain period of time are fixed, which is not in line with most transaction scenarios. 
Current performance optimization is mainly based on blockchain storage optimization, blockchain redesign, and reinforcement learning to improve performance implementation. In order to meet the challenges of blockchain systems in IIoT, J. Huang et al. [99] proposed a directed acyclic graph structured blockchain system based on the IIoT consensus mechanism of credit. An adaptive PoW algorithm for power-constrained iot devices is used to reduce power consumption in the consensus mechanism and improve system throughput by utilizing its asynchronous consensus model, which can adjust the difficulty of the PoW based on the behavior of the nodes, reducing the difficulty for honest nodes and increasing the difficulty for malicious nodes. They also proposed an access control scheme based on symmetric cryptography in a transparent blockchain system, providing users with a flexible approach to data rights management. In terms of storage optimization, there are metadata-based blockchain research and lightweight blockchain research.
\subsection{Consensus algorithm}
Blockchain technology is essentially a peer-to-peer distributed database that maintains the consistency of the blockchain system through consensus algorithms, realizes trust between different nodes and calculates the corresponding rights and interests of each node. POW consensus algorithm is a proof-of-work consensus mechanism and one of the core technologies of Bitcoin. However, with the continuous innovation of blockchain technology, the defects of POW consensus algorithm in performance and security are more and more obvious, and improved protocols based on POW consensus algorithm are constantly proposed. In the POW Consensus protocol, new coin rewards and transaction fees protect the security of the Bitcoin network. If a greedy saboteur is able to concentrate more computing power than an honest node, it launches a 51\% attack. From an economic perspective, the benefits of following the rules of the Bitcoin system are in most cases greater than the benefits of an attack. The shortcomings of POW consensus algorithm are mainly in the following three aspects.

Waste of resources. Mining requires a lot of hash operations, requires electricity and various computing resources, and the hash values found do not actually have any practical use value.

The network performance is low. Because the POW consensus algorithm limits the time of bitcoin block to 10 minutes, transaction confirmation takes at least 10 minutes, and currently only supports 7 transactions per second, which is not suitable for high-concurrency commercial applications.

PoW consensus algorithm computing power centralization. At present, the mining pool is the main force, and it is basically impossible for individual miners to survive, and the mining pool with high computing power has the option, which leads to the concentration of computing power.

The core contradiction of POW algorithm is block size and block interval. Increasing block capacity can improve throughput, but too many blocks will cause network congestion, increase the time and efficiency of inter-node consensus, and may reduce block efficiency. Reducing the outgoing block interval can also increase throughput, but the shortening of the outgoing block interval will cause more frequent chain forks and increase security issues such as double flowers.
Therefore, with the development of public chain consensus mechanism, POW consensus algorithm has produced many variants. There are two ways to improve its performance and security. One is to transform the growth mode of the chain, redistribute accounting rights, and reduce disorderly competition and block interval without changing the core of POW proof-of-work. One is not to modify the content of the POW consensus algorithm, and control the transaction volume on the chain through the off-chain expansion mechanism to improve the efficiency of the blockchain.
\subsection{Privacy protection}
The blockchain-based distributed ledger integrates a variety of technologies such as asymmetric encryption systems, P2P networks, consensus algorithms, and smart contracts to ensure the consistency and imtamability of transaction records. However, the ledger sharing mechanism in blockchain technology also brings privacy threats, and the privacy protection of user identity, account address, transaction content and other information has become the focus of research. In the actual use of the blockchain system, in order to ensure the traceable, verifiable and other characteristics of the recorded data on the blockchain, all data must be disclosed to all nodes in the blockchain network. This feature, while ensuring security and verifiability, allows malicious attackers to directly access the data recorded in the blockchain ledger and snoop on user privacy through analysis of the data. By analyzing the transaction data recorded in the blockchain ledger, the attacker discovers the rules, associates the different addresses and transaction data of the user, and further corresponds to the real identity of the user. Such attacks mainly rely on address clustering and identity information mining. In order to prevent the disclosure of personal privacy, the existing research is similar to the method of network privacy protection, which realizes information concealment and concealment by mixing personal information with other users' information.

In order to resist the ledger analysis technology, according to the assumptions based on the technology, the researchers propose a defense mechanism for exchanging assets and confusing addresses, namely, address obfuscation mechanism. Different users exchange assets with each other through transactions, so as to achieve the effect of confusing user addresses and protecting the privacy of each user. Because the address obfuscation mechanism is carried out through the exchange of assets, it is usually called the mixing mechanism, and the transaction used to exchange assets is called the mixing transaction.
\section{Possible future trends}
Blockchain has shown its potential in both industry and academia. In this section, we discuss possible future directions from three aspects: blockchain testing, artificial intelligence, and blockchain applications.
\subsection{Blockchain testing}
Many different types of blockchains have emerged recently, and as of now, coindesk has more than 700 cryptocurrencies listed on it. When incorporating blockchain into their business, users must select the type of blockchain to meet their business requirements. Therefore, there is a need for a recognized and universally applicable blockchain testing mechanism to test different blockchains. At the same time, blockchain performance is an important indicator of blockchain application, and blockchain testing is also conducive to objectively evaluating the stability, scalability, security and other performance indicators of a blockchain.

Blockchain testing can be divided into three phases: selection phase, performance testing phase and operation and maintenance testing phase. In the selection stage, the type of blockchain and its matching with the business scenario are mainly tested. The performance test stage mainly tests different performance indicators of blockchain, which is mainly divided into single indicator test and business scenario indicator test. The single indicator test only tests the blockchain performance indicators, while the business scenario indicator test needs to test the important performance indicators in the scenario combined with the specific scenario. The operation and maintenance test phase mainly tests the convenience and stability of blockchain operation and maintenance.
\subsection{Artificial intelligence}
The combined application of AI and blockchain technology is still in the exploratory stage, but from the point of view of reinforcement learning methods being used to optimize blockchain performance, AI technology is bound to be able to assist blockchain. As a law and regulation on the blockchain, the essence of smart contract is to reach a contract call through a transaction, so it can be seen that its essence is not a smart contract. And AI technology helps build intelligent prophecy machines so that smart contracts can become smarter.
\subsection{Blockchain applications}
Most blockchain applications are currently in the financial sector, and more and more applications in different fields are emerging. In this paper, the application of blockchain in the traditional industries of intelligent transportation, smart grid and Internet of Things is reviewed. In the future, blockchain will definitely be applied to more fields to support the development of the field or enhance the system in the field, and the combination of blockchain and reputation mechanism makes the possibility of malicious nodes to do evil greatly reduced, which is conducive to the more general promotion of blockchain to autonomous systems and other fields.
\section{Conclusion}
Blockchain has shows its potential to transform traditional industries with its key characteristics such as decentralization, persistence, anonymity, and auditability. As an innovative technology, blockchain technology has now become the cornerstone of the field of information technology and has also had a huge impact on many academic research directions. In this paper, we present a systematic and comprehensive overview of blockchain. We begin with an overview of blockchain technology, including blockchain architecture and key features of blockchain. Then the research and application of blockchain in various fields are analyzed and summarized from different angles. In addition, we enumerate some of the challenges and issues that hinder the development of blockchain, and summarize some of the existing approaches to address them. Some possible development directions in the future are put forward. Today, based on the application of blockchain applications continue to emerge, and we plan to conduct in-depth investigations on blockchain-based applications in the future based on the principle of technology for the benefit of mankind.

\section*{Acknowledgments}
This work described in this paper was supported by the National Natural Science Foundation of China under Grant No. 61862063, 61502413, 61262025; the National Social Science Foundation of China under Grant No. 18BJL104; the Science Foundation of Young and Middle-aged Academic and Technical Leaders of Yunnan under Grant No. 202205AC160040; the Science Foundation of Yunnan Jinzhi Expert Workstation under Grant No. 202205AF150006; the Open Foundation of Yunnan Key Laboratory of Software Engineering under Grant No. 2020SE301; the Science Foundation of "Knowledge-driven intelligent software engineering innovation team".The authors would like to thank Prof.Xuan Zhang for his constructive comments.

At the same time, we also want to thank Dr.Liu Jinzhuo, Software School of Yunnan University, for his guidance in writing this article. In addition, I also want to thank Fan Qiyuan, Yu Hao, Zhao Haiyang for their efforts in the work of this paper.

\begin{IEEEbiography}[{\includegraphics[width=1in,height=1.25in,clip,keepaspectratio]{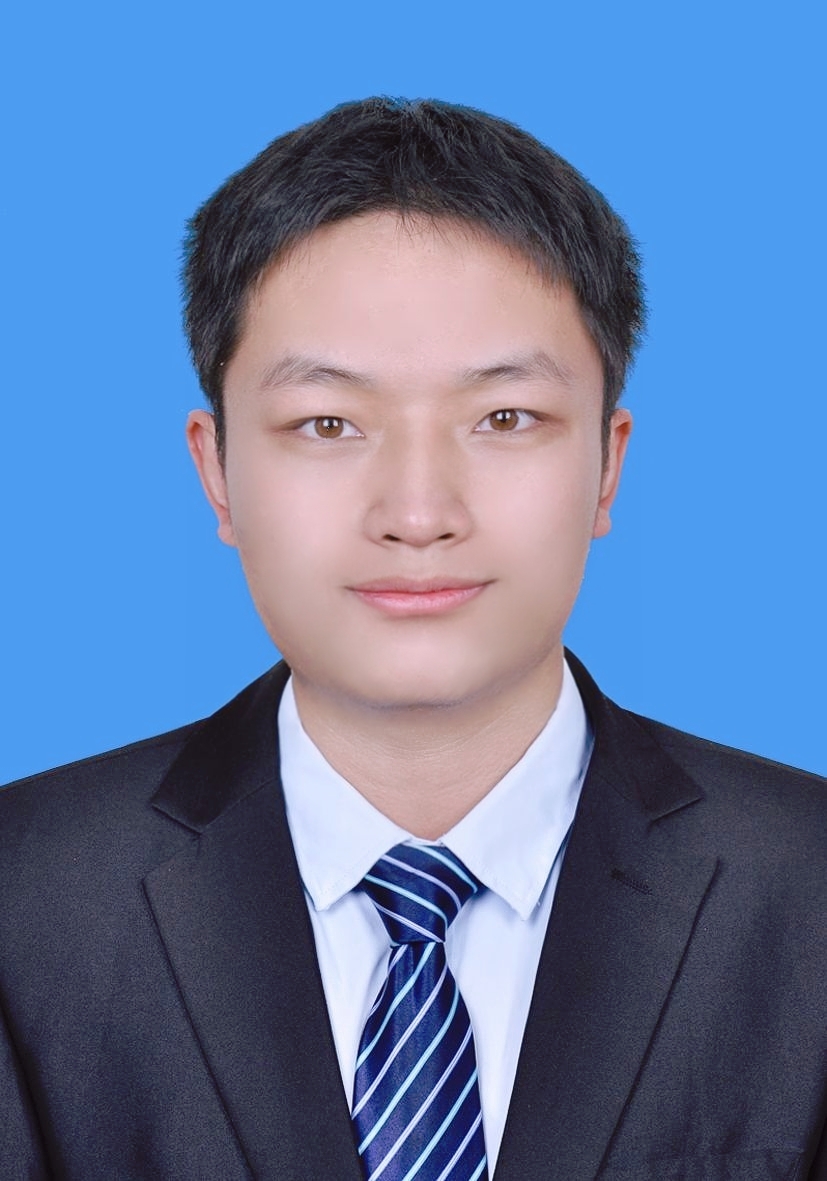}}]{Min An} (Graduate Student Member, IEEE) received the B.S. degree in computer science and technology from Northwest Minzu University, Lanzhou, China, in 2022. 

He is currently pursuing the M.S. degree with the Yunnan Key Laboratory of Software Engineering, School of Software, Yunnan University, Kunming, China. His current research interests include reinforcement learning, time series forecasting, blockchain technology and federated learning.
\end{IEEEbiography}
\begin{IEEEbiography}[{\includegraphics[width=1in,height=1.25in,clip,keepaspectratio]{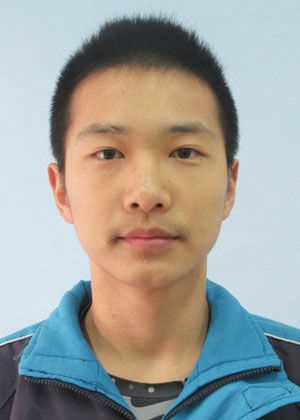}}]{Qiyuan Fan} received his Bachelor's degree in Wuhan University of Technology of School of Computer Science, China in 2022.

He is currently studying for a master's degree in the School of Software, Yunnan University, Kunming, China. His current research interests include tensor decomposition.
\end{IEEEbiography}
\begin{IEEEbiography}[{\includegraphics[width=1in,height=1.25in,clip,keepaspectratio]{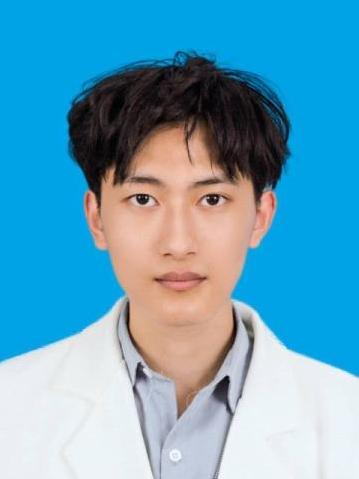}}]{Hao Yu} received the B.S. degree in computer science and technology from Southwest University of Science and Technology, Chendu, China, in 2022. 

He is currently studying for a master's degree in the School of Software, Yunnan University, Kunming, China. His current research interests include computer vision.
\end{IEEEbiography}
\begin{IEEEbiography}[{\includegraphics[width=1in,height=1.25in,clip,keepaspectratio]{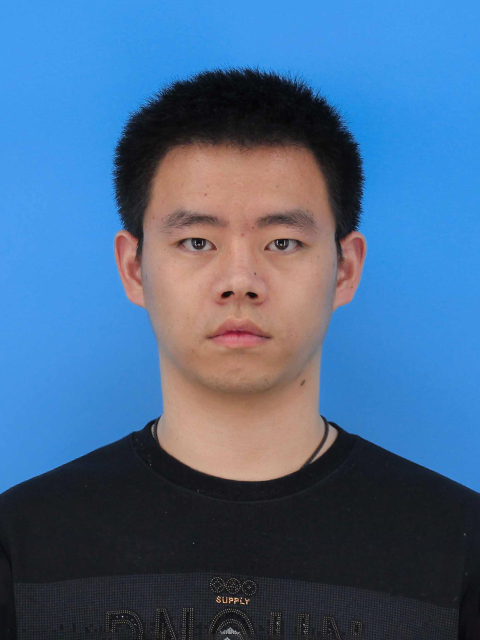}}]{Haiyang Zhao} received his Bachelor's degree in Geotechnical and Urban Underground from Hunan University of Science and Technology, China in 2022.

He is currently studying for a master's degree in the School of Software, Yunnan University, Kunming, China. His current research interests include computer vision and autonomous driving.
\end{IEEEbiography}

\vfill

\end{document}